# Ion transport through differently charged nanoporous membranes: from a single nanopore to multi-nanopores

Hongwen Zhang,[1,2,3] Bowen Ai,[1,2,3] Zekun Gong,[1,2,3] Tianyi Sui,[4] Zuzanna S. Siwy,[5] and Yinghua Qiu[1,2,3*]

1. Key Laboratory of High Efficiency and Clean Mechanical Manufacture of Ministry of Education, National Demonstration Center for Experimental Mechanical Engineering Education, School of Mechanical Engineering, Shandong University, Jinan, 250061, China

2. State Key Laboratory of Advanced Equipment and Technology for Metal Forming, Shandong University, Jinan 250061, China

3. Shenzhen Research Institute of Shandong University, Shenzhen, 518000, China

4. School of Mechanical Engineering, Tianjin University, Tianjin, 300072, China

5. Department of Physics and Astronomy, University of California, Irvine, California 92697, United States

*Corresponding author: yinghua.qiu@sdu.edu.cn

**Abstract**

Nanoporous membranes, leveraging their high-throughput characteristics, have been widely applied in fields such as molecular separation and energy conversion. Due to interpore interactions, besides the applied voltage and solution environment, the ion transport properties in porous membranes are influenced by the pore number and spacing. Here, to understand and control the transport properties of nanopore arrays, we systematically investigate the ion transport characteristics through membranes with different charge properties, pore numbers, and interpore distances. Using numerical simulations, we analyzed local ionic concentrations and electric potential in nanopore arrays containing nanopores with uniformly charged walls as well as unipolar diodes i.e. pores containing a junction between a charged zone and a neutral zone, and showed significant ion concentration polarization (ICP) for all studied cases. As the number of pores increased and the interpore spacing decreased, the enhanced interpore interactions through ICP led to a greater deviation of the total ionic current from the linear superposition of single-pore currents. Conversely, in bipolar nanopores whose walls contain a junction between positively and negatively charged zones ICP becomes negligible, and interpore interactions are substantially reduced. Furthermore, for membranes with various charge properties, the total current through nanopore arrays presents different quantitative dependence on the pore number under varying pore spacings. Our findings clarify the mechanism of interpore interactions in modulating ion transport through porous membranes, providing critical insights for designing nanofluidic devices based on nanopore arrays, such as nanopore-array sensors.

**Keywords:**

**TOC**

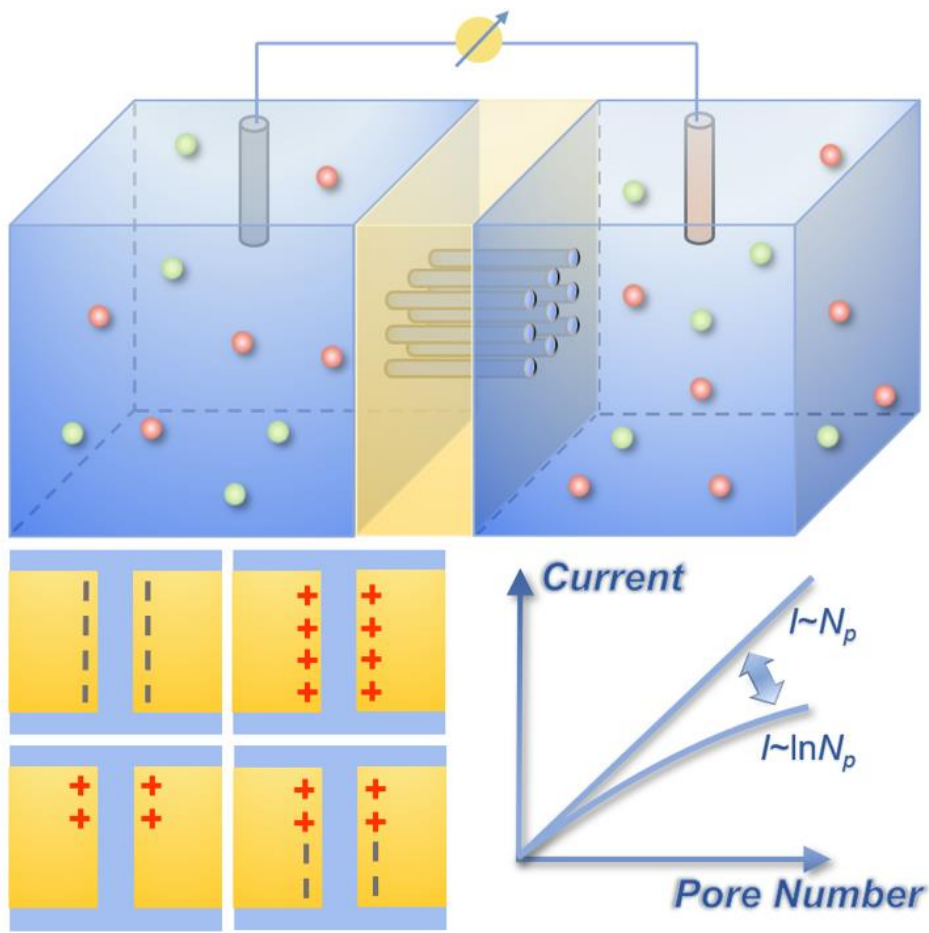

## Introduction

Nanopores provide a versatile platform to investigate the transport of ions and molecules in nano-confined spaces,[1, 2] and have already found applications in various fields, such as material separation,[3, 4] mass transport,[5, 6] energy conversion,[7, 8] and nanofluidic sensing.[9, 10] Under such highly-confined pores, the charge properties at solid-liquid interfaces provide significant modulation of the characteristics of ion transport.[2] Because of the strong electrostatic interactions between surface charges and free ions, counterions accumulate near the charged surfaces to form electric double layers (EDLs),[11] which render nanopores ion selectivity[12] and promote rapid transport of counterions through the nanopore.[6] Research on single nanopores, considered as the functional element of porous membranes, has provided significant insight into transport in confined geometries.[13] However, many practical applications including seawater desalination,[3, 14] osmotic energy conversion,[15, 16] and electroosmotic pumps,[17, 18] among others[9, 19] require high fluxes that can be offered only if a membrane is highly porous.

Studies have shown that the characteristics of ion transport in porous films can be significantly different from those in single nanopores.[20, 21] The strategy of describing ion transport in porous membranes by linearly superimposing the characteristics of ion transport through individual nanopores is no longer accurate which results from the communication among the individual nanopores. Gadaleta et al.[20] investigated the current characteristics through low-aspect ratio nanopore arrays of varying pore numbers and pore spacing fabricated on silicon nitride chips. Their results indicate that for a 1D linearly arranged nanopore array, the total conductivity $G_{total}$ follows a relationship of $G_{total} \sim N_p G_S / \log N_p$, in which $N_p$ is the number of nanopores and $G_S$ is the conductivity through a single nanopore. While for the 2D arrangement of nanopores, the conductivity through the nanopore array follows $G \sim \sqrt{N} G_S$. Their sub-additive ion current through nanopore arrays results from overlapping access resistance zones. Enhanced ion concentration polarization (ICP)[22] across the porous membrane can lead to significant suppression of the current. Using microfluidic chips containing nanochannel arrays, Ahmed et al.[23] dynamically monitored the evolution of ion depletion zones near the nanochannels by the fluorescence imaging technique, including the formation, expansion, and overlap of ion depletion zones. Combining the imaging with electrochemical recordings, they demonstrated that the overlap of ion depletion zones in nanopore arrays is closely related to the much lower current through the porous membrane than the expected value from single-pore superposition. Green et al.[24] designed nanochannel arrays with varying spacing and numbers to investigate the interaction between channels and their effects on ion transport. Their results highlight the importance of properly controlling channel spacing to enhance the current performance. To investigate the influence of the pore spacing on the pore-pore interaction, Liu et al.[25] fabricated nanopore arrays on SiN membranes with varying ratios of the pore spacing $r_c$ to the pore radius $r_s$. When $r_c/r_s$ was small, the diffusion zones of adjacent nanopores overlapped significantly, causing the deviation of the actual current from the theoretical value. At $r_c/r_s = 56$ or higher, individual pores in the array could work independently without the interaction from neighboring pores.

The dependence of ion current through a nanopore on the number and arrangement of neighboring pores can however be advantageous in some cases. In a striking example, an array consisting of nanofluidic channels where neighboring nanopores carried opposite surface charges exhibited enhanced transport compared to arrays with only negatively or positively charged channels.[26] Interactions between nanopores in an array were also found to improve nanopore sensing. In one example, Hu et al.[27] showed that due to a profile of an electric field created across an array of four closely located pores, the speed of DNA translocation can be significantly reduced without compromising the signal-to-noise ratio. Drndic et al.[28] reported a 3D array of pores connected in parallel and series and demonstrated that overlapped electric fields enable the detection of short DNA molecules. In many other cases, however, nanopore arrays consisting of chemically similar pores exhibit decreased ionic conductance and selectivity, resulting in diminished performance of membranes.[29, 30]

Research on ion transport through porous membranes not only elucidates the transport characteristics of ions in practical applications but also provides useful guidance for the design of porous membranes. Nanofluidic experiments provide the actual transport behaviors of porous membranes, typically representing the average current through the pores. While simulations can provide the microscopic details of ion transport in individual nanopores on porous films. As an example, 3D numerical models provided details of local ionic concentrations and ion transport in arrays composed of 3, 6, and 9 nanopores was simulated with different pore spacings.[31] The models revealed the overlapping depletion zones due to ion concentration polarization that are responsible for current suppression in arrays consisting of homogeneously charged pores.[29] Additionally, pore parameters such as pore length, pore diameter, and surface charge density were also found to affect the ion transport through nanopore arrays.[30, 32, 33] The higher surface charge densities in nanopore arrays can lead to an expansion of the ion depletion zone and enhance the suppression of ion transport.[34]

Due to the high computational cost required by 3D simulations, existing 3D studies often employ small models, use coarse grids, or ignore the fluid flow in the system. 2D simulations can be applied for large and ultra-fine models, which have been widely applied in the study of porous membranes. 2D models[35] for example revealed proton concentration distributions in individual nanopores in an array. The position-dependent proton concentration was shown to lead to different surface charge densities on each pore wall, which in turn affected ion transport and fluid flow in the multipore systems.

With a dual-nanopore array, Raza et al.[36] investigated the variation in ion current at different nanopore spacing and pore diameters. Their results demonstrated that larger pore diameters require a greater spacing for independent functioning.

The inter-pore interaction within nanopore arrays significantly affects ion transport properties, particularly at high pore densities, where ICP leads to deviations of the current from the ideal value predicted by single-pore superposition. Optimizing the performance of nanopore arrays critically depends on the proper control of pore spacing and pore density.[37] However, there remains a lack of systematic research on the quantitative description of ionic currents and the underlying ion transport mechanisms in nanopore arrays with different charge distributions under the electric field. Here, 2D simulations were conducted to investigate the ion and fluid transport through nanoporous membranes. For the nanopore arrays, we considered individual nanopores with uniform, unipolar, and bipolar surface charges. From the obtained microscopic details of fluid flow and the ionic current characteristics of individual pores on the porous membrane, which are challenging to obtain directly through experiments, we explored the interaction mechanisms between nanopores and achieved a quantitative description of the current in porous membranes. The study reveals that when a nanoporous membrane carries a single type of charge (e.g, uniformly negative or unipolar charge in a diode), membranes with a higher number of pores and smaller pore spacing exhibit severe ICP at the membrane-solution interfaces, leading to pronounced interactions between individual pores. Consequently, due to the weaker interaction between nanopores as the pore spacing increases, the dependence of the total current ($I_{total}$) on pore number ($N_p$) transitions from a logarithmic ($I_{total} \sim \ln N_p$) to a linear ($I_{total} \sim N_p$) relationship. In contrast, when the membrane carries both positive and negative charges (e.g., alternating charged or bipolar), ICP is effectively alleviated. The weakened interpore interactions result in a total current that more closely follows a linear proportionality to the pore number.

**Simulation Details**

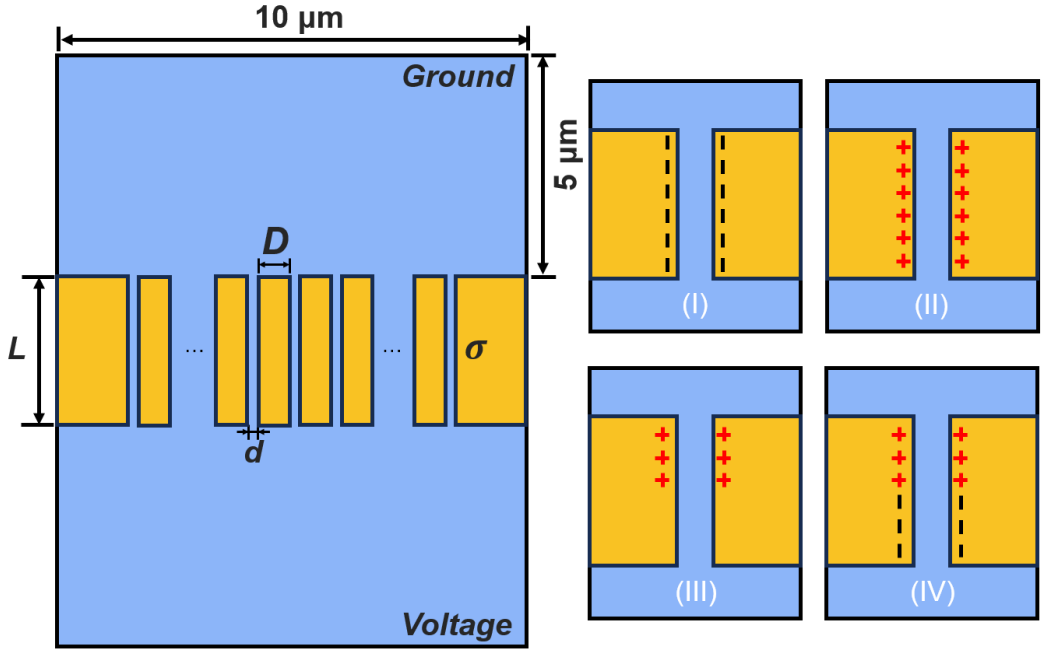

Figure 1. Scheme of a nanopore array that was considered by finite element simulations. The nanoporous membrane is located between two reservoirs with 5 μm in width and 10 μm in length. A voltage is applied across the membrane. The inset shows the four differently charged unit nanopores considered in this work. These pores are negatively charged (I), positively charged (II), unipolarly charged in a diode (III), and bipolarly charged (IV). Nanopores are 10 nm in diameter and 1000 nm in length.

With COMSOL Multiphysics, two-dimensional nanofluidic simulations were conducted to investigate ion transport through nanoporous membranes, which can shed light on how transport through a pore depends on the presence, number, and spacing of neighboring pores.[35, 38] As shown in Figure 1, the porous film is located between two reservoirs. Both reservoirs are 5 μm in width and 10 μm in length. The nanopore number $N_p$ in the film varies from 1 to 10. The diameter $d$ and length $L$ of nanopores on the porous film are 10 and 1000 nm, respectively. The separation of individual nanopores is defined as the distance between two neighbor pore boundaries, denoted as $D$, which is used to consider various porous films with different pore densities. Here, the nanopore separation varies from 5 to 1000 nm, corresponding to $1 \times 10^8 \sim 4.5 \times 10^{11}$ pores/cm$^2$.[21, 39]

Considering the practical surface charge properties in nanofluidic porous materials, four types of nanopores were considered with different charge properties. As shown in Figure 1, these pores are negatively charged (I), positively charged (II), unipolarly charged known as unipolar diodes (III), and bipolarly charged, known as bipolar diodes (IV). For unipolar and bipolar diodes, the junction between differently charged regions on pore walls is located in the middle of the pore.[40] The surface charge density was set to ±0.08 C/m$^2$ for positively and negatively charged surfaces.[41, 42] Aqueous solutions of 100 mM KCl were applied in the system with the dielectric constant of water 80. The diffusion coefficients of K$^+$ and Cl$^-$ were $1.96 \times 10^{-9}$ m$^2$/s and $2.03 \times 10^{-9}$ m$^2$/s.[43] The system temperature was maintained at 298 K. A voltage was applied at the boundary of both reservoirs, varying from −1 to 1 V, with a default value of 1 V.

In our simulations, coupled Poisson-Nernst-Planck and Navier-Stokes equations (Eqs. 1-4) were solved to comprehensively consider the ion distribution in the electric double layers (EDLs) near charged pore walls, ion transport, and fluid flow in the nanopores and reservoirs.[6, 44-47]

$$\varepsilon \nabla^2 \varphi = -\sum_{i=1}^{N} z_i F C_i \tag{1}$$

$$\nabla \cdot \mathbf{J}_i = \nabla \cdot \left( C_i \mathbf{u} - D_i \nabla C_i - \frac{F z_i C_i D_i}{RT} \nabla \varphi \right) = 0 \tag{2}$$

$$\mu \nabla^2 \mathbf{u} - \nabla p - \sum_{i=1}^{N} \left( z_i F C_i \right) \nabla \varphi = 0 \tag{3}$$

$$\nabla \cdot \mathbf{u} = 0 \tag{4}$$

where $z_i$, $J_i$, $C_i$, and $D_i$ are the valence, ionic flux, concentration, and diffusion coefficient of ionic species $i$ (cations or anions). $\varepsilon$ is the dielectric constant of solutions, and $\boldsymbol{u}$ is the fluid velocity. $\varphi$, $N$, $F$, $R$, $T$, $p$, and $\mu$ are the electric potential, number of ion types, Faraday constant, gas constant, temperature, pressure, and liquid viscosity, respectively.

The ion currents through porous films were obtained by integrating the total ion flux at the reservoir boundary with Eq.(5).[5, 6]

$$I = \int_S F \left( \sum_i^2 z_i \mathbf{J}_i \right) \cdot \mathbf{n} \; \mathrm{d}S \tag{5}$$

where $\mathbf{n}$ is the unit normal vector, and $S$ represents the boundary of the reservoir, respectively.

A similar mesh strategy to our previous works was applied (Figure S1).[5, 6] The ultrafine mesh size of 0.1 nm was set on inner pore walls to consider the ion distribution within the EDLs and the microscopic characteristics of ion and fluid transport through nanopores. 0.1 nm mesh was also used on the exterior surface within 3 μm from the pore wall to the reservoir boundary, as well as between neighboring nanopores to investigate the microscopic characteristics near nanopore orifices. For the other parts on exterior surfaces, a 0.5 nm mesh size was selected to improve the calculation efficiency without affecting the calculation accuracy.

## Results and Discussion

***Arrays with uniformly charged nanopores exhibit suppressed ion current due to ion concentration polarization.***

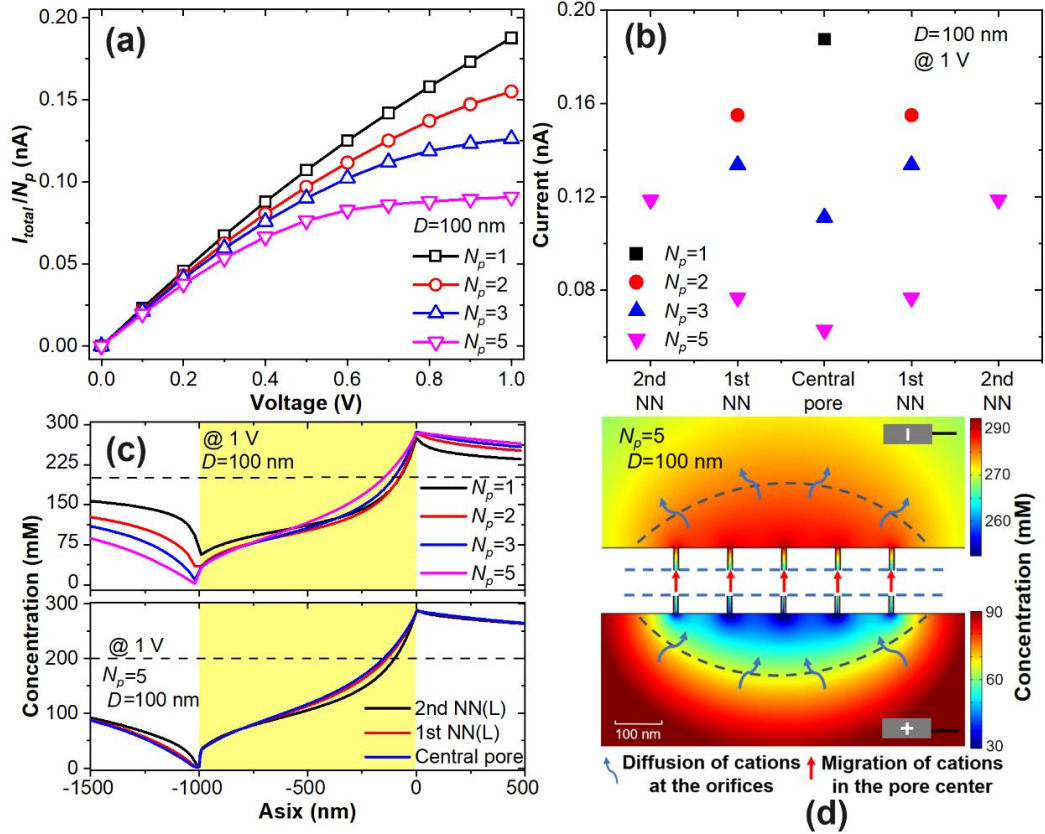

Figure 2. Ionic transport properties of arrays consisting of uniformly negatively charged nanopores. (a) Current-voltage (I-V) curves of arrays with different numbers of pores. The pore separation was set to 100 nm, corresponding to a pore density of $1 \times 10^{10}$ pores/cm$^2$. (b) Ionic current through each unit nanopore at 1 V. (c) Axial distributions of the total ion concentration in the nanopore and at pore entrances. The upper figure shows the averaged total concentration distribution through nanopores in the porous membranes with 1 to 5 nanopores. The lower figure presents the concentration distribution through each unit nanopore collected on the 5-pore membrane. (d) Distributions of ion concentration at $N_p$=5.

At solid-liquid interfaces, solid surfaces usually carry charges that originate from various mechanisms, such as the ionization of surface chemical groups.[48] For example, polymer porous membranes fabricated by the track-etching method[49, 50] and silicon nitride porous films prepared by focused ion or electron beams[51] are negatively charged at neutral pH due to deprotonation of carboxyl and silanol groups, respectively. Figure 2 shows the characteristics of ion transport through a uniformly negatively charged nanoporous film, with the consideration of the nanopore number $N_p$=1, 2, 3, and 5. Individual nanopores are 10 nm in diameter and 1000 nm in length. The interpore distance was set to 100 nm. Corresponding current-voltage (I-V) curves are shown in Figure 2a. In addition, to quantitatively compare the ion current through porous membranes with different nanopore numbers, the average current through each pore on the film is calculated by dividing the total current $I_{total}$ by the pore number $N_p$ (Figure 2b).

In the case with a single pore, the I-V curve is linear at applied voltages less than 0.3 V. When the voltage exceeds 0.3 V, the effects of ICP become significant, resulting in the characteristic S-shape of the I-V curves with the appearance of limiting current.[6, 52] As the pore number increases from 2 to 5, the current decrease becomes more pronounced. In the system with 2 nanopores, the normalized current of a single pore in an array at 1 V is ~82% of that through the single-pore case. When the nanopore number increases to 5, the normalized ion current decreases to ~48% of the value of a single independently functioning nanopore (Figure S2). In addition, as the pore number increases, the threshold voltage that marks the onset of the limiting current decreases. The occurrence of the limiting current through nanopores is due to the ICP with the formation of a depletion zone at the pore entrance and part of the pore volume. The depleted regions are responsible for the decreased conductance. As the nanopore number increases, the phenomenon of ICP at both pore ends becomes more pronounced (Figures 2c and 2d).[34, 39]

We also consider ion current through each pore array consisting of 3 and 5 nanopores. The individual pores are labeled as the central pore and the nth neighbor nanopore (the nth NN). For the case with 2 nanopores, both nanopores are the 1st NN.

By integrating the ion flux at the nanopore entrance, the ion current through each unit nanopore in the porous film can be obtained. As reported before,[31] and shown in Figure 2b, with the increase of the pore number, the ion current through individual nanopores significantly decreases and exhibits position dependence. Nanopores located closer to the array center have lower currents, and those closer to the array boundary exhibit greater currents. For $N_p$ = 3 and 5, the current in the central pore decreases to ~59% and ~34% of the single pore current value, respectively (Figure S3). As $N_p$ increases to 5, the current value of the central nanopore is ~40% of that of the edge nanopore.

The limiting current through nanopores and the position-dependent current in unit nanopores are due to ICP-induced depletion zones at the pore entrances. On one hand, therefore, porous membranes containing many pores can increase transport, however, ICP limits the transport of individual pores in a membrane. As the number of pores increases, the number of ions that can diffuse from the bulk to the pore orifice decreases, resulting in more significant ion depletion at the pore entrance.[30, 33] This corresponds to the lower ion current through individual nanopores that have more adjacent nanopores (Figure 2d). Finally, by comparing the simulation results with and without the consideration of the Navier-Stokes equation in the model,[53] EOF is confirmed to induce much more significant ICP across porous membranes, especially highly packed membranes (Figure S4).

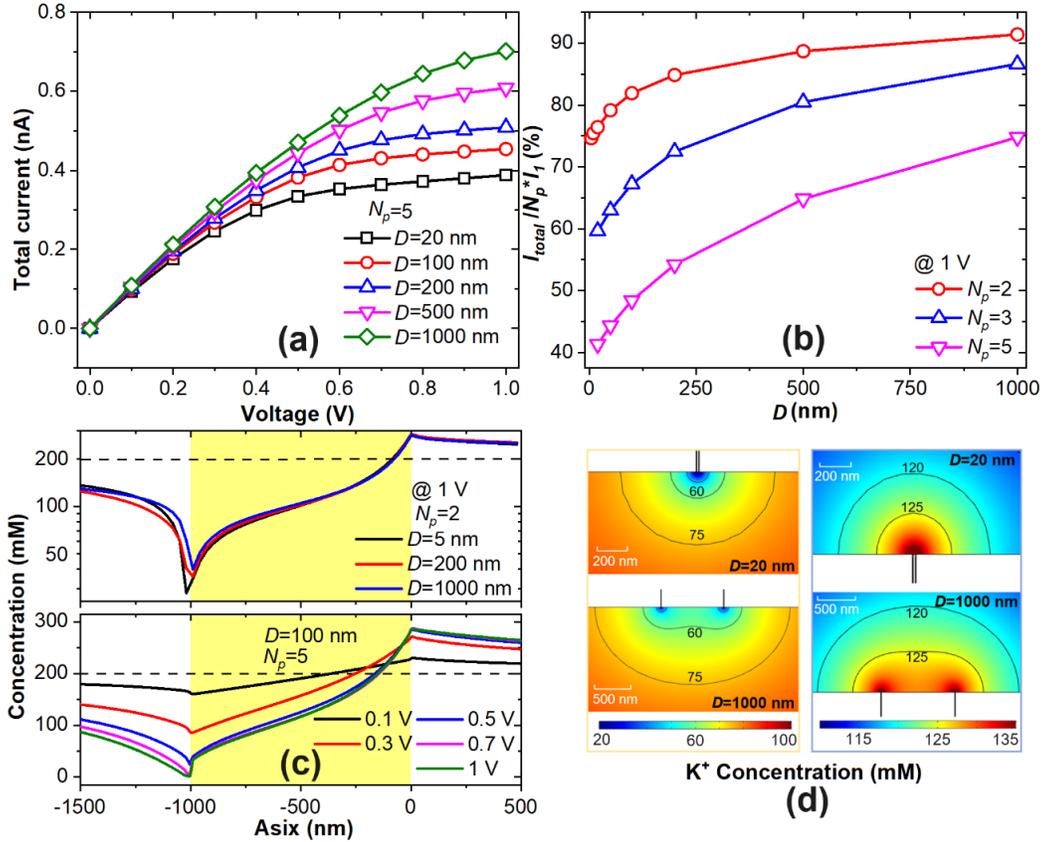

Figure 3. Ionic transport through arrays with uniformly charged pores at different pore separations. (a) I-V curves and (b) relative average current of membranes consisting of 2-5 pores with different pore separations. All current values are normalized to the current obtained through the single-pore membrane. (c) Axial distributions of the average total ion concentration inside the nanopore. The upper figure presents the average concentration distribution collected on the two-pore membrane with different pore separations. The lower figure shows the average concentration distribution inside nanopores on the porous membranes at different voltages. (d) 2D concentration distribution of $K^+$ ions at the entrance and exit of the nanopores ($N_p$=2).

The pore density in porous membranes is determined by the pore separation. To understand how interpore distance in membranes affects ion transport, we considered arrays of $N_p$=2, 3, and 5 with a pore separation in the range between 20 nm and 1000 nm. Figure 3a shows the I-V curves obtained with a 5-pore membrane with different pore separations. As the pore separation changes from 1000 to 20 nm, the limiting current appears at ~1 V at $D$=1000 nm, and at ~0.5 V at $D$=20 nm. At the transmembrane potential of 1 V, the current values through the membrane decrease as well, from ~0.71 to ~0.39 nA, by ~45%. The changes are mainly due to the enhanced ICP at the pore entrances in more closely packed pores. Similar trends also occur in cases with porous membranes containing 2 or 3 nanopores, where for the same pore separation, the average current through the porous membrane has a larger decrease as the pore number increases (Figure 3b).

The axial distributions of average total ion concentration at different voltages are also investigated to reveal the origin of the more obvious limiting current in I-V curves at $N_p$=5 (Figure 3c, lower panel). For $D$=100 nm, as the applied voltage increases from 0.1 to 0.5 V, the stronger electric field enhances the migration speed of ions, resulting in the linear I-V, although the degree of ICP inside the nanopore is noticeable. At 0.7 V, the more significant ICP results in the formation of a depletion zone at the pore entrance, which limits the current. As the voltage approaches 1 V, saturated ion enrichment and depletion are created at the respective pore entrance, and the transmembrane current continues to be limited. As expected, considering the interactions between adjacent nanopores in the porous film, the current through nanopores under high voltages exhibits a decreasing trend as the pore separation further decreases.

Figure 3d provides the two-dimensional distributions of $K^+$ ion concentration across the porous membrane with two nanopores at separations of 20 and 1000 nm. The left and right figures show the local ion depletion and enrichment at the entrance and exit of two nanopores, respectively. We selected a depleted concentration of 60 mM for indication. At $D$=1000 nm, the $K^+$ ion concentration of 60 mM appears at a distance of ~250 nm from the pore orifice. As the two pores approach, the ion depletion regions at the pore entrance of each nanopore overlap. At $D$=20 nm, the K$^+$ ion concentration of 60 mM appears at a much closer distance of ~150 nm to the pore entrance, meaning a more significant ion depletion and corresponding to a stronger interaction between nanopores. At the exit of the nanopore, a large number of ions need to migrate away from the nanopore to the bulk. The limited pore area at a small separation between nanopores slows down the ions' diffusion from the near-pore area to the bulk solution.[6] Therefore, at the pore separation of 20 nm, a more significant ion enrichment happens compared to that at a separation of 1000 nm (Figure 3d, right panel).

***Arrays with alternating distribution of positively and negatively charged closely packed nanopores offer enhanced transport compared to a set of independently functioning nanopores.***

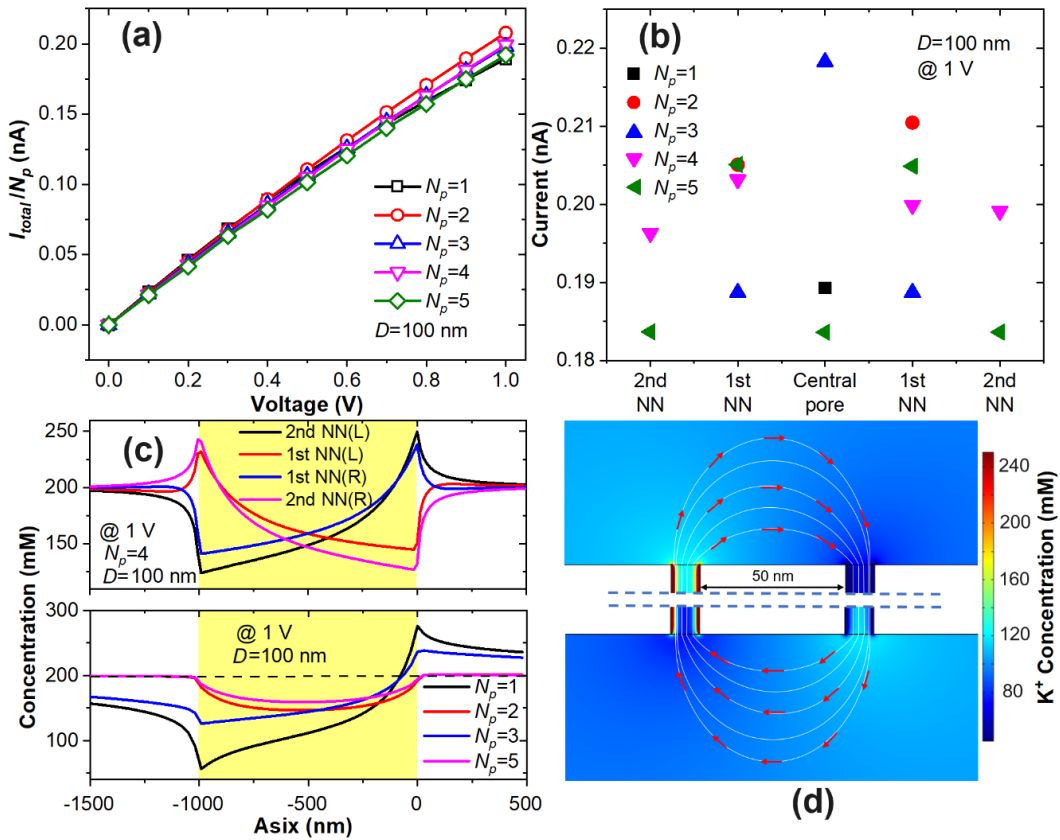

Figure 4. Electrochemical properties of nanopore arrays containing an alternating

distribution of positively and negatively charged nanopores. (a) I-V curves obtained using membranes with different pore numbers. The pore separation was set to 100 nm. (b) Ionic current through each unit nanopore at 1 V. (c) Axial distributions of the ion concentration inside nanopores. The upper figure presents the concentration distributions inside each unit nanopore on a four-pore membrane. The lower figure shows the average total ion concentration distribution inside nanopores on the porous membranes with 1 to 4 nanopores. (d) 2D concentration distribution of $K^+$ ions at the entrance and exit of the nanopores ($N_p$=2).

The majority of previous work on arrays was performed with systems containing identical pores with the same chemistry and functionality,[54, 55] for example, all negatively charged pores, or all pores containing ionic diodes. Recently, an experimental and modeling report was published that showed the advantages of membranes containing alternating nanopores with positive and negative surface charges.[26] Namely, it was predicted that such membranes exhibit higher conductance than even a set of independently working homogeneously charged pores. We therefore considered a similar system containing nanopores with positive and negative surface charges, as shown in Figure 1 (I-II) to understand the mechanism of the current enhancement. Figure 4a shows the I-V curves through porous films with various pore numbers of $N_p$=1, 2, 3, 4, and 5. The arrangement of surface charges in the constituent nanopores is shown in Figure S5. As proposed before, the alternating distribution of positive and negative nanopores can indeed eliminate the ICP across the porous membrane effectively.[26] No obvious limiting current appears in the I-V curves (Figure 4a). The average current through each porous membrane with different pore numbers shares almost the same values.

To understand the current enhancement as well as how the pore location affects its transport, Figure 4b shows the current through individual nanopores in arrays with different numbers of pores and different interpore distances. Figure S6 provides the normalized currents of individual pores in porous films divided by the current through the membrane with a single nanopore. At $N_p$ of 2 and 4, the current value of each pore has an increase of ~5% and ~10% compared to that in the case of a single nanopore. In our simulations, the application of KCl solutions leads to similar ion currents through positively and negatively charged nanopores because of the close diffusion coefficients of $K^+$ and $Cl^-$ ions. The current values of positively charged nanopores are slightly higher than those of negatively charged ones.

It is worth noting that at an odd pore number of 3, the current in the central pore is significantly larger than that through the nearby nanopores, ~115% of the current in the case with a single nanopore. While the neighboring nanopores (1st NN) have roughly equal current values to those through the single-pore membrane. A 3-pore array with two negatively charged nanopores accompanying one positively charged pore is also considered (Figure S7). Although the arrangement of differently charged nanopores is reversed, the current of the central pore is still greater than that of the adjacent nanopores.

The enhancement of ionic conductance of arrays with alternating pores with positive and negative surface charges, stems from the depletion zone at neighboring pores to be located on the opposite sites of the membrane. In this case, due to the opposite ICP across differently charged nanopores, the ICP inside the central nanopore is effectively eliminated, and abundant free ions can be supplied by the nearby nanopores. Figure 4c upper panel, reveals that the ions enriched in the unit nanopores can provide additional charge carriers for adjacent nanopores, which thereby promote the current inside the pores and suppress the generation of ICP. Figure 4c shows the axial distributions of ion concentration in a porous film with alternating positive and negative charges. The concentration is the average total ion concentration in each unit pore. Figure 4c lower panel, shows that the inhibition of ICP in porous films with alternating positive and negative charges exhibits pore-number dependence, meaning that this inhibition is only effective when the number of nanopores is even. When the number of nanopores on the porous film is odd, the suppression effect of ICP is weak due to the imbalance between ion transport and supply, resulting in a slight increase in current. In this case, the central nanopore experiences a more significant increase in current due to the replenishment of ions from both sides of the nanopore. Following a similar strategy, a simulation containing a 5-pore membrane with alternating positively and negatively charged nanopores was conducted (Figure 4b). In this case with 3 negative and 2 positive nanopores, the unit nanopores present two current levels. The center pore and boundary pores have a lower current. The 1st NNs have a larger current.

In these porous films with alternating positively and negatively charged nanopores, due to the mutual influence between adjacent nanopores, a flow loop of fluid and ions can be generated (Figure 4d). This fluid loop can promote the transport of ions into nanopores, which is enhanced as two nanopores approach each other due to the strengthened interaction between nanopores. The average ion concentration inside the pores is increased, which results in a higher current. From the additional simulations without the consideration of the fluid flow (Figure S8), a weak current decrease appears even with the same alternating distribution of positive and negative pores. Our modeling provides evidence that the ion transport contributed by the convection plays an important role in the enhancement of current through porous membranes.[53]

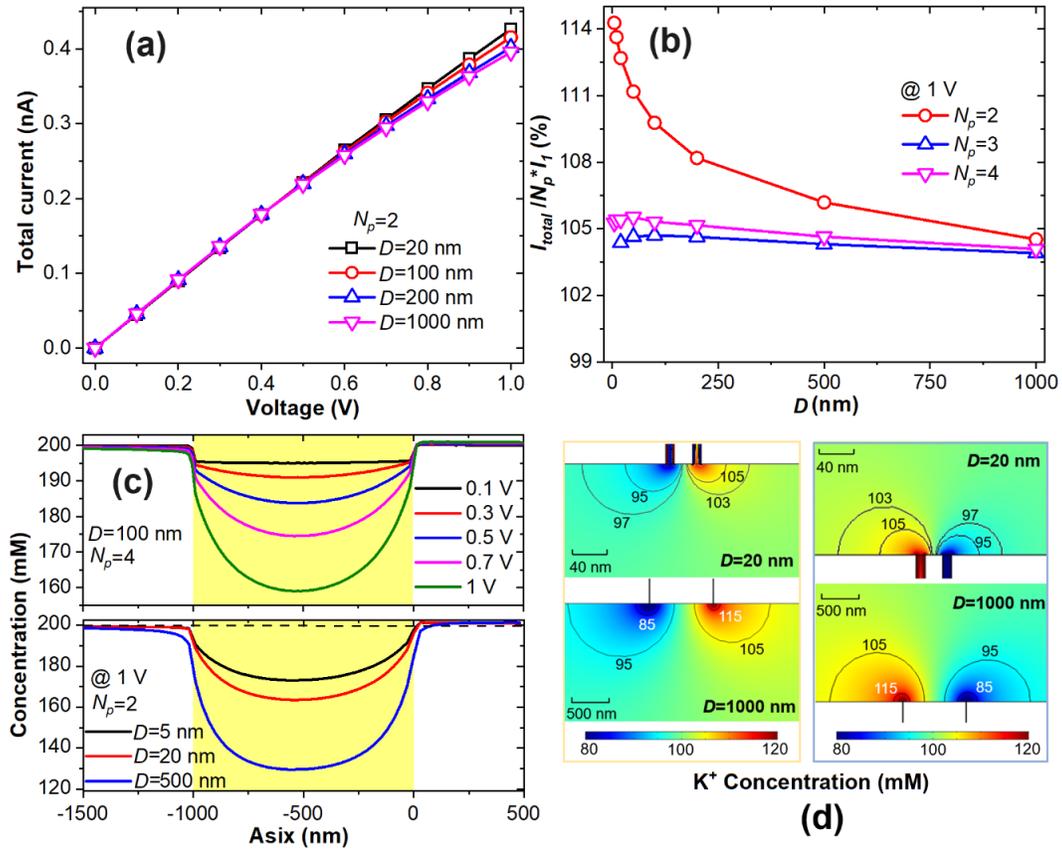

Figure 5. Transport properties of nanopore arrays consisting of nanopores with positive and negative surface charges. (a) I-V curves obtained using membranes with different pore separations. The pore number was set to 2. (b) Relative average current through multi-pore membranes with different pore separations. All current values are normalized to the current obtained with the single-pore membrane. (c) Axial distributions of the average total ion concentration inside nanopores. The upper figure shows the average concentration distribution inside nanopores in porous membranes consisting of four pores at different voltages. The lower figure presents the average concentration distribution calculated for a two-pore membrane with different pore separations. (d) 2D concentration distribution of $K^+$ ions at the entrance and exit of the nanopores ($N_p$=2).

For porous films with alternating positive and negative nanopores at $N_p$=2, 3, and 4, the influence of the pore separation on current is also explored. Figure 5a shows the I-V curves obtained with two-pore membranes at different pore separations. For uniformly charged porous films, a smaller pore separation can induce greater ICP and current suppression. In contrast, in the case of porous films with alternating positive and negative nanopores, a smaller pore spacing suppresses the ICP and enhances current (Figure 5d). As the pore separation decreases, i.e. the pore density increases, the total current through the porous film presents an increasing trend. For example, when the pore separation decreases from 1000 to 20 nm, the current increases from ~0.4 to ~0.43 nA by ~7% in a two-pore membrane at 1 V. We believe that this current increase is related to the increased ionic concentration in the pore for smaller interpore distances (Figure 5c, lower panel).

Figure 5b shows the normalized current value through porous membranes relative to the current value $I_1$ in the case with a single pore membrane. The current through the porous membrane with alternating positive and negative nanopores exhibits a different trend compared to that through a uniformly charged porous membrane, i.e. the average current in the nanopores becomes higher than that through the membrane with a single nanopore, and no limiting current appears at stronger voltages. For a two-pore membrane at a small pore separation of 5 nm, the average current value is ~115% of that through the single-pore membrane, Figure 5b. As the pore separation increases, although the promoting effect on current weakens, there is still a 4% increase in current at $D$=1000 nm. Figure S9 shows a cross-sectional view of the concentration distribution in the case with 5 pores at $D$=1000 nm, indicating that at a large separation, nanopores function independently. At $N_p$=3 and 4, the promoting effect of alternating positive and negative nanopores on current decreases with the variation of pore separation. The difference in the current increase between $D$=20 and 1000 nm is less than 2%.

***Arrays with unipolarly charged nanopores exhibit suppressed ion current due to strong ion concentration polarization.***

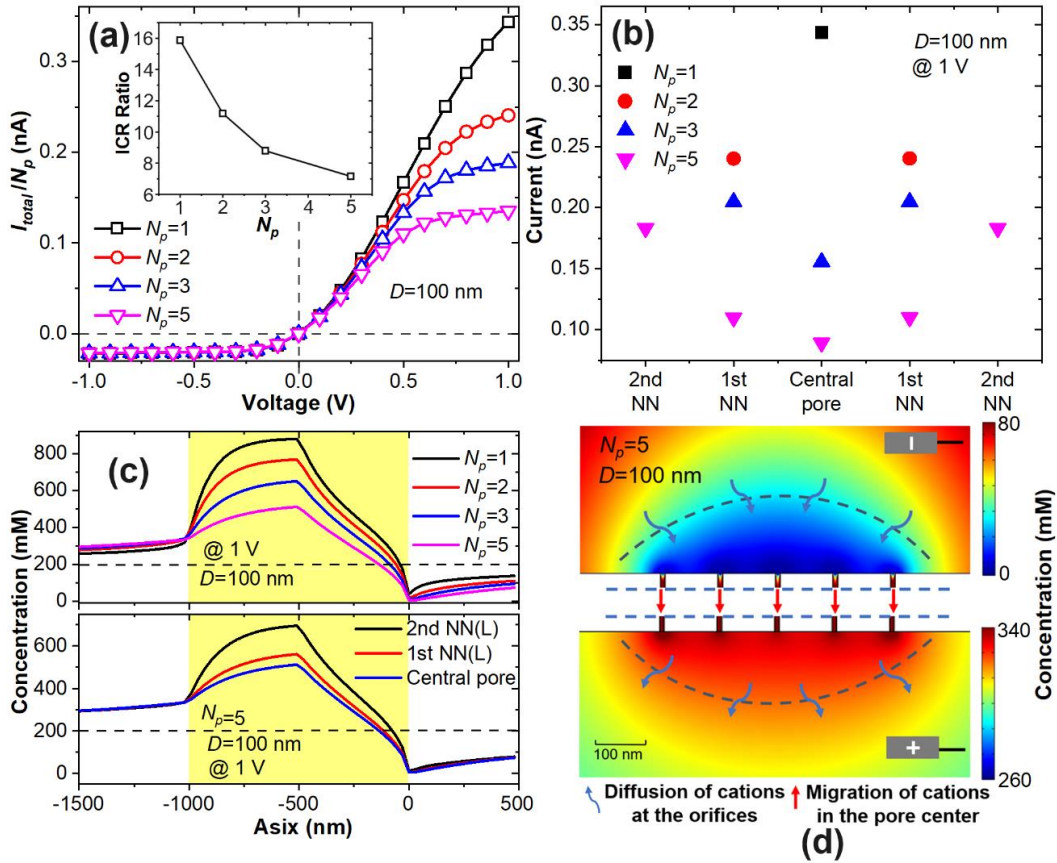

Figure 6. Ion transport through arrays consisting of ionic unipolar diodes. (a) I-V curves obtained using membranes with different nanopore numbers. The inset shows the ICR Ratio with different nanopore numbers. (b) The ionic current through individual nanopores at 1 V. (c) Axial distributions of the ion concentration through nanopores. The upper figure shows the average total ion concentration distribution through nanopores on the porous membranes with 1 to 5 nanopores. The lower figure presents the concentration distribution inside individual nanopores in the 5-pore membrane. (d) Distributions of ion concentration at $N_p$=5.

As the next step, we have considered nanopore arrays whose nanopores exhibited broken electrochemical symmetry and rectified ion current.[56, 57] We began the analysis with unipolar ionic diodes i.e. nanopores containing a junction between a charged zone, in our case positively charged, and a neutral zone, Figure 1 (III-IV). As reported before,[56] I-V curves of such pores are asymmetric with currents for voltages of one polarity higher than currents for voltages of the opposite polarity. The larger and smaller currents correspond to the "on" and "off" states of the rectified nanopores. These nanopores with the preferred direction of ion transport may have applications in the field of energy conversion. [58, 59]

Figure 6a shows I-V curves obtained for thin membranes with different numbers of unipolar diodes ($N_p$=1, 2, 3, and 5), where the two zones, charged and uncharged, had equal lengths. In our electrode configuration shown in Figure 6d, the rectifiers exhibit their "on" and "off" states under positive and negative voltages, respectively. To characterize the degree of ionic current rectification (ICR), the ICR ratio is used, which is denoted by ICR Ratio = $|I_{+1 \text{ V}}|$ / $|I_{-1 \text{ V}}|$.[60] From the inset of Figure 6a, the ICR ratio of the unipolar membrane with a single nanopore can reach ~16. I-V curves are also obtained with porous membranes of multi-nanopores with a separation of 100 nm. With $N_p$ varying from 2 to 5, nanopores at the "off" state present almost the same current values. While for the current at "on" state, the current exhibits number-dependent and voltage-dependent trends.

At different pore numbers, and 100 nm interpore distance, varying degrees of rectification appear. The ICR ratios at $N_p$=2, 3, and 5 are ~11, ~9, and ~6, which decrease by ~30%, ~45%, and ~60%, from that of a single nanopore, respectively. In order to explain the dependence, the axial distribution of ion concentration at positive voltages is shown in Figure 6c. With the unchanged ion concentration under negative voltages (Figure S10), the ICR ratio presents a decreasing trend as the nanopore number increases due to the lower ionic enrichment inside nanopores (Figure 6c, upper panel).

I-V curves of nanopore arrays with even 2 pores exhibit the presence of limiting current at positive voltages which limits the rectification properties of the system. With $N_p$=2, as the voltage increases from 0 to ~0.8 V, an approximately linear I-V is observed through the porous membrane. When the applied voltage exceeds 0.8 V, a limiting current starts to appear that is absent for a single unipolar diode membrane. Similar to cases with uniformly charged nanopores shown above, as the nanopore number increases to 3 and 5, limiting current appears at lower trans-membrane potentials due to overlapping depletion zones of neighboring pores, created by ICP (Figure 6d).

Figure 6b shows the ion current through each unipolar diode in arrays with 2~5 diodes. Similar to what we observed with a uniformly charged porous membrane, ion current through a single diode in an array decreases as the number of pores increases and is position-dependent. Figure S11 shows the normalized distribution of current values of unit nanopores on unipolar porous membranes with different $N_p$. For the array with two unipolar diodes, the current value through each pore is ~70% of that of a single-pore membrane. Compared to the cases with uniformly charged nanopores, the current drop of ionic unipolar diodes in an array is greater than that through uniformly charged porous membranes. At $N_p$=3, boundary nanopores present a larger current than the central pore. As $N_p$ increases to 5, the current difference between boundary nanopores and the central pore becomes more significant, because the ion enrichment inside boundary nanopores is much higher than the central pore (Figure 6c, lower panel). In this case, the current of the central pore is only ~26% of that through the single-pore membrane.

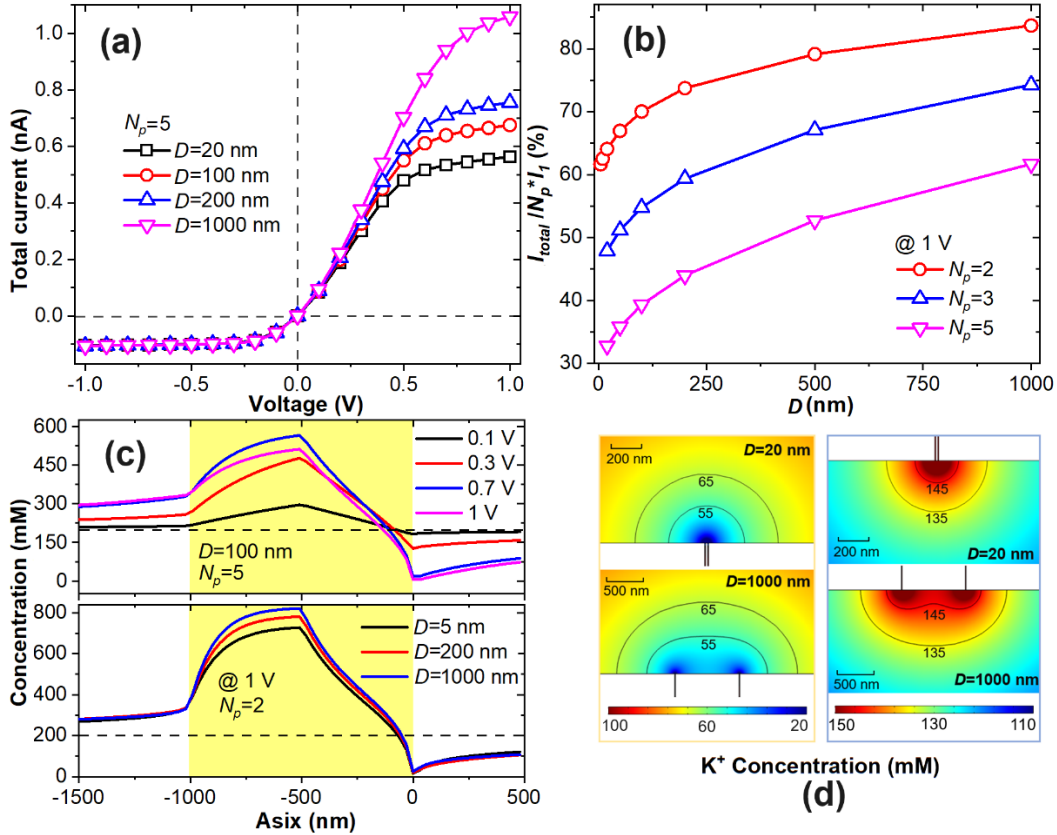

Figure 7. Transport properties of nanopore arrays consisting of unipolar diodes as a function of pore number and interpore distance. (a) I-V curves obtained using membranes with different pore separations ($N_p$=5). (b) Relative average current through multi-pore membranes with different pore separations. All current values are normalized to the current obtained through the single-pore membrane. (c) Axial distributions of the average total ion concentration inside the nanopore. The upper figure shows distributions of the average concentration through nanopores on the porous membranes at different voltages. The lower figure is the concentration distribution collected at different pore separations. (d) 2D concentration distribution of K$^+$ ions at the entrance and exit of the nanopores ($N_p$=2).

The influence of the pore separation on the ion transport through unipolar porous membranes with different nanopore numbers is also considered. Figure 7a shows the I-V curves through a unipolar 5-pore membrane at different pore separations. In a denser nanopore array, the phenomenon of limiting current becomes more obvious. At 1 V, as the pore separation shrinks from 1000 to 20 nm, the current decreases from ~1.06 to ~0.57 nA, by ~46%. According to Figure 7c lower panel, in an array with a larger pore separation between adjacent nanopores, more ions can be enriched inside the nanopores, which results in a higher current. The relationship between ICR ratios and the pore separation in the unipolar porous membrane is also investigated, as shown in Figure S12. For a 2-pore membrane, the ICR ratio is ~9.9 at $D$=5 nm, which increases to ~13.3 as the pore separation expands to 1000 nm. This can be attributed to the more obvious ion enrichment inside nanopores resulting from the weaker interaction between adjacent nanopores in a sparser array (Figure 7c, lower panel). Figure 7b shows the normalized current through individual nanopores in porous membranes with different pore numbers. For the two-pore membrane at $D$=5 nm, the current under positive voltage is ~62% of that through the single-pore membrane, which is significantly lower than the normalized values of ~84% at D=1000 nm. For different pore numbers, the normalized current values share a similar trend with the pore separation. Compared to the cases with multiple uniformly charged nanopores, at the same pore separations, unipolar diodes exhibit a stronger suppression of the current through individual nanopores (Figure 7d and Figure S13). For nanopores that function as unipolar diodes, positive voltages induce enhancement of ionic concentration inside the pore. This enhancement is higher than in uniformly charged pores, and consequently, a lower partial drop of the transmembrane voltage occurs inside the unipolar pore, which in turn induces a higher partial voltage at the pore mouth. This stronger electric field at the pore entrance promotes ion migration and forms a more obvious ICP.

Under negative voltages, the effect of pore separation on the ICP can be ignored (Figure S14), because the transport is limited not by ICP at the pore entrance but rather by the depletion zone created in the pore due to the junction between charged and neutral zones of the pore walls. Based on the current enhancement at positive voltages induced by weaker ICP, as the pore separation increases, the ICR ratio presents an increasing trend.

***Arrays with bipolarly charged nanopores reduce ICP and eliminate limiting current behavior.***

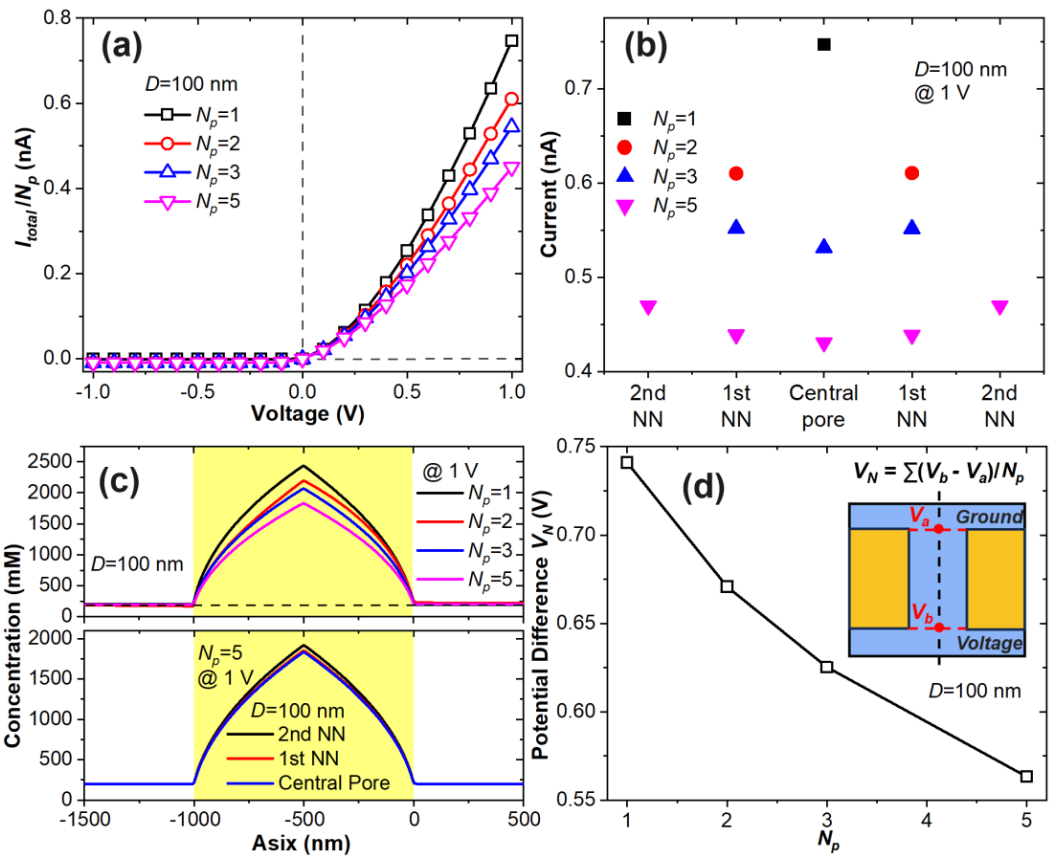

Figure 8. Ion transport in arrays consisting of bipolar diodes. (a) I-V curves obtained using membranes with different nanopore numbers. (b) Ionic current through each unit nanopore of arrays at 1 V. (c) Axial distributions of the ion concentration through the nanopore. The upper figure shows the averaged total ion concentration with different $N_p$. The lower figure presents the concentration distribution inside individual nanopores in the 5-pore membrane. (d) The potential difference between $V_a$ and $V_b$ in the porous membranes, $V_a$ and $V_b$ is the potential value at the central points of the pore entrance and exit.

The second nanopore system with asymmetric surface charges is the bipolar nanopore where a junction is created between a zone with positive surface charges and a zone with negative surface charges. According to our previous work,[40] bipolar nanopores can produce significant current rectification, mainly because both oppositely charged parts of the nanopore are selective to the anions and cations in the solution. Transport properties of ionic bipolar diodes are similar to the properties of a semiconductor p-n junction where the zone with negative (positive) surface charges is an ionic equivalent of a p-doped (n-doped) semiconductor. At the electrode configuration shown in Figure S15, under positive and negative voltages, the ion enrichment and depletion can be induced inside the bipolar nanopores, corresponding to the high and low current values in the I-V curves, respectively.

As the nanopore number increases, the concentration through the nanopores gradually decreases under positive voltages, the degree of ion enrichment weakens, and the current decreases (Figure 8c, upper panel). To further investigate the fluid flow and pore-pore interactions in bipolar porous membranes, the effective potential difference across the nanopore (Figure 8d) was obtained by measuring the potential distribution inside the nanopores (Figure S16). The potential in the porous membrane is obtained along the pore axis of all individual pores.[6, 51] For each nanopore, the potential difference is obtained by subtracting $V_a$ from $V_b$ (inset of Figure 8d). $V_a$ and $V_b$ are the potentials at the center point of the pore entrance and exit, respectively. For porous membranes, the potential difference is the average potential difference over all individual nanopores, i.e. $\sum (V_b - V_a)/N_p$. For the single pore membrane, the potential difference across the nanopore is ~0.74 V. In the cases at $N_p$ = 3 and 5, the potential difference across the nanopore decreases to ~0.63 and 0.56 V, respectively, by ~15% and 24%. Figure S17 shows the effective potential difference at the exit and entrance of each unit pore in the bipolar porous membrane. The obtained potential difference across individual nanopores has the same trend as the current values. Due to the decreased potential difference across the nanoporous membrane, weaker electric fields result in a lower degree of ion enrichment inside nanopores, corresponding to a smaller average current through the membrane. Please note that with other simulation models, the potential difference across porous membranes of various pore numbers was also explored, as shown in Figure S18. In the other three cases considered above, the potential difference across porous membranes also exhibits a similar decreasing trend with the increase of the nanopore number.

Figure 8b shows the current of individual nanopores in bipolarly charged porous membranes. With multi-nanopores in the membrane, the current through each nanopore becomes lower than that of the single-pore membrane. At $N_p$=5, the current in the central pore is only ~58% of the current through the single pore film. Different from the current behavior of unipolar nanopores, the current values of individual bipolar nanopores share similar values in the pore arrays, which have no obvious position dependence. At $N_p$=5, the current difference between the center pore and the boundary pore is ~5% (Figure S19). Individual nanopores share a similar degree of ion enrichment (Figure 8c, lower panel). Please note that at the entrance and exit of the bipolar nanopores, no obvious ICP occurs (Figure S20).

The increase in the nanopore number can result in a decrease in the potential difference across nanopores. Nanopores with lower potential differences can induce weaker electric field strength, resulting in a smaller current through the pore. No limiting current appears in the cases with bipolar membranes, even at 2 V (Figure S21).

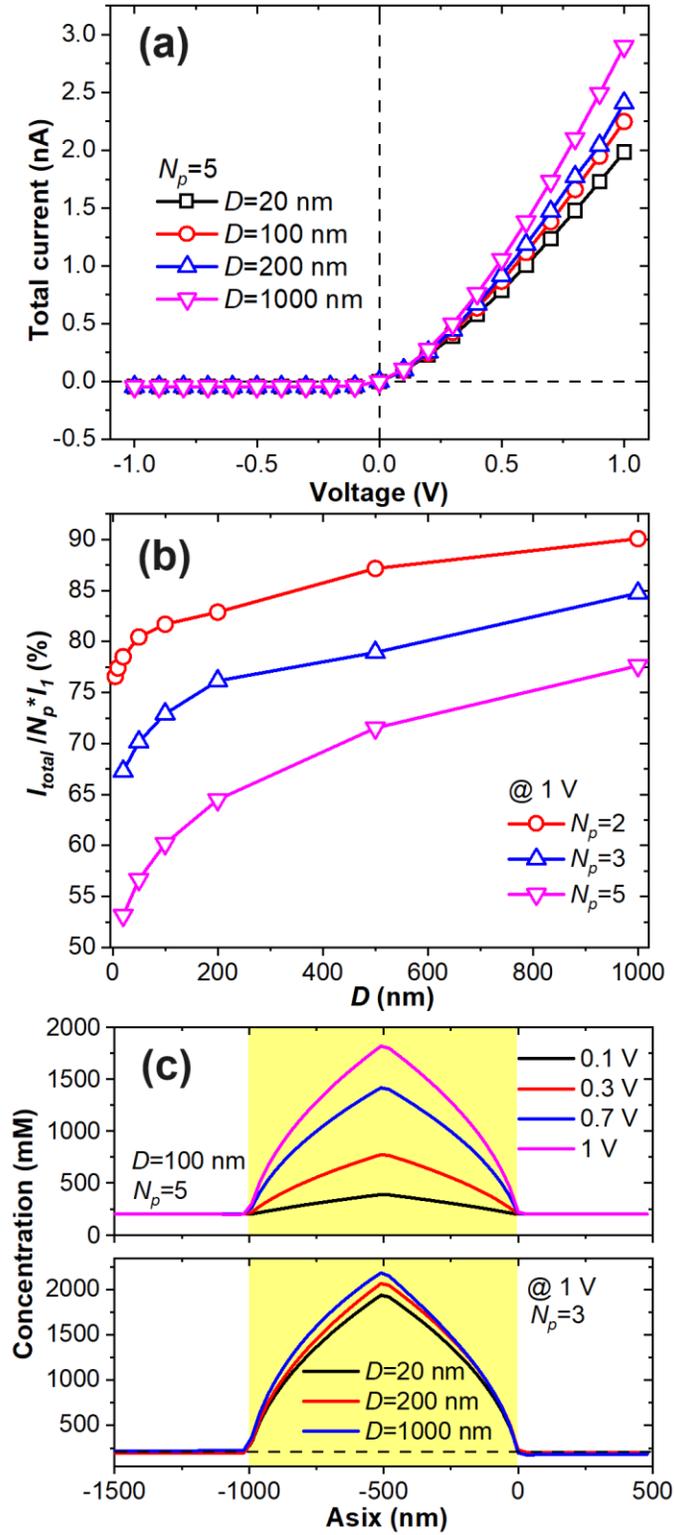

Figure 9. Ion current and local ionic concentrations in arrays consisting of bipolar diodes under different pore separations. (a) I-V curves obtained using membranes with different

pore separations ($N_p$=5). (b) Relative average current through multi-pore membranes with different pore separations. All current values are normalized to the current obtained through the single-pore membrane. (c) Axial distributions of the average total ion concentration through the nanopore. The upper figure shows distributions of the average concentration through nanopores on the porous membranes at different voltages. The lower figure is the concentration distribution collected at different pore separations.

Figure 9a shows the I-V curves for different values of pore separation, *D*. As the voltage increases, the larger degree of ion enrichment inside the nanopore leads to a higher current (Figure 9c, lower panel), such that no limiting current is observed. For bipolar single pores, the current can reach ~0.75 nA at 1 V, around twice the current value through single unipolar diodes. Due to the much smaller current at negative voltage than unipolar nanopores, an ICR ratio of ~169 can be achieved (Figure S22). However, as *D* decreases, due to the mutual influence between nanopores, a decrease in current value occurs at positive voltages. Therefore, the ICR ratio of porous membranes is lower than that of the single-pore membrane.

The effect of the pore proximity on the array conductance is quantified in Figure 9b in more detail. As the pore separation increases from 20 to 1000 nm through the 5-pore membrane, the current value at 1 V increases from 2 to 2.89 nA, by ~45%. For porous membranes with a constant nanopore number, as the pore separation increases, the interaction between pores weakens and the current at positive voltages gradually increases. For a two-pore membrane, at *D*=20 nm, the current at 1 V is ~1.17 nA. When the distance between the pores increases to 1000 nm, the current value is ~1.35 nA, an increase of ~15%. At $N_p$=3 and 5, the current at 1 V increases by ~26% and ~46%, respectively. As the pore separation increases, the ion enrichment inside nanopores becomes more significant at positive voltages, resulting in a higher current and an increase in the ICR ratio (Figure 9c). An array with more nanopores has a more obvious increasing trend in the ICR ratio with the pore separation.

***Quantitative relationship between ion current and the pore number in four cases of arrays with different surface charge distributions***

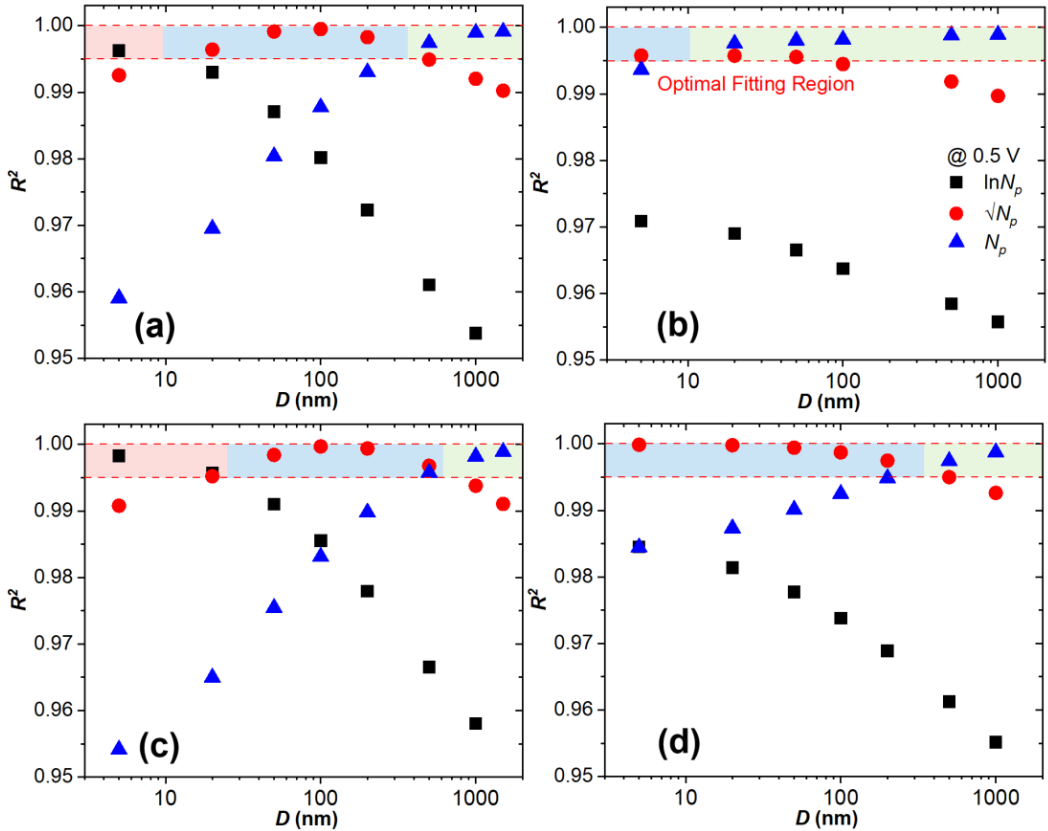

Figure 10. Coefficient of determination $R^2$ with different numerical fitting equations under different pore spacings ($D$) at 0.5 V. The pore numbers varied from 1 to 10. (a) Arrays with uniformly charged nanopores. (b) Arrays with alternating distribution of positively and negatively charged nanopores. (c) Arrays with unipolar diodes. (d) Arrays with bipolar nanopores.

Finally, we have considered the possibility of identifying a quantitative relationship between the total current $I_{total}$ and the number of pores $N_p$ at different pore spacings for the four types of porous membranes described here.[20] We began the process by fitting the current through nanopore arrays as a function of the number of pores under various voltages and interpore distances. Three possible dependences of current on the number of pores were considered: $I_{total} \sim \ln N_p$, $I_{total} \sim \sqrt{N_p}$ and $I_{total} \sim N_p$. As an example, Figure S23a shows that for arrays with uniformly charged nanopores at a pore spacing of 5 nm, due to the strong interaction between adjacent pores, the fitting of current versus $I_{total} \sim \ln N_p$ performs better than $I_{total} \sim \sqrt{N_p}$ or $I_{total} \sim N_p$. To evaluate the performance of the prediction in the current, the coefficient of determination $R^2$ was applied, which is calculated with Equation 6.[61, 62]

$$R^2 = 1 - \frac{\sum \left(I_i - \hat{I}_{fit}\right)^2}{\sum \left(I_i - \bar{I}\right)^2} \tag{6}$$

where $I_i$ represents the actual current value, $\hat{I}_{fit}$ denotes the fitted value, and $\bar{I}$ is the average of the actual current values. Note that a value of $R^2$ approaching 1 indicates a better agreement between the theoretical prediction and our simulation data.

Figure 10 shows the values of $R^2$ at 0.5 V for each pore spacing for arrays with homogeneously charged pores, arrays with alternating positively and negatively charged pores as well as unipolar and bipolar diodes. Different colors in Figure 10 indicate the optimal fitting equation under varying pore spacings. For arrays with uniformly charged nanopores, when the pore spacing is above ~10 nm, compared to $\ln N_p$, $\sqrt{N_p}$ provides the best fit among the three considered functions (Figure S23b). This is attributed to the weakened interaction between nanopores at a larger pore spacing. As the pore spacing exceeds 250 nm, $I_{total}$ follows a linear growth relationship with $N_p$ (Figure S23c), i.e. the nanopores effectively function as independent entities. Please note that we don't attempt to provide an accurate theoretical prediction for the current through porous membranes.

From the literature, the relationship between the current and the number of pores might depend on the concentration of salt, pore length, diameter, and applied voltage.[32, 37] Here, based on our results, with the ICP becoming more significant, the dependence of current on the pore number changes from $I_{total} \sim N_p$ to $I_{total} \sim \ln N_p$ gradually.

Due to the alternating distribution of positive and negative nanopores, ICP is effectively suppressed across the membrane, which induces a linear relationship between the current and the pore number $N_p$, when the pore spacing is greater than 10 nm (Figure 10b). The equation $I_{total} \sim \sqrt{N_p}$ fits better when the pore spacing is less than 10 nm (Figure S23d).

In arrays of unipolar diodes, the current and pore number relationship follows a similar trend as observed in negatively charged porous membranes (Figure 10c). As pore spacing decreases, the dependence of current on pore number transitions from $I_{total} \sim N_p$ to $I_{total} \sim \ln N_p$. At small pore spacings ($D$=5 nm), the current curves under different voltages are proportional to $\ln N_p$ at various pore numbers (Figure S24a) due to the strong interaction between pores. When the pore spacing surpasses 25 nm, the interaction between nanopores weakens, and the fitting equation $I_{total} \sim \sqrt{N_p}$ yields a higher $R^2$ value closer to 1 (Figure S24b). As the pore spacing further exceeds 600 nm, the nanopores operate independently, and $I_{total}$ exhibits a linear dependence on $N_p$ (Figure S24c).

In bipolar porous membranes, ICP at various pore spacings is much weaker than uniformly or unipolarly charged nanopores. While considering the interaction between nanopores, under a pore spacing smaller than ~250 nm, the current through porous membrane follows $I_{total} \sim \sqrt{N_p}$, which has much larger $R^2$ values than the $\ln N_p$ fitting (Figures S24d-e). As the pore spacing increases to above 250 nm, the interaction between nanopores diminishes, and $I_{total}$ exhibits a linear growth relationship with $N_p$ (Figure S24f). From Figure 10d, due to the weaker interaction between nanopores in bipolar membranes, the current prediction with $I_{total} \sim \sqrt{N_p}$ and $I_{total} \sim N_p$ has a higher accuracy.

**Conclusions**

Our simulations reveal the mechanisms by which pore number and spacing influence ion transport in nanoporous membranes with distinct surface charge properties. Specifically, for uniformly charged membranes, densely arranged nanopores exhibit a significant reduction in total current due to ICP. The induced ICP is particularly pronounced near the central pores, leading to position-dependent and number-dependent ionic currents at individual pores. Additionally, the unipolar membrane induces severe ICP at pore entrances, further suppressing the ion flux. In contrast, nanoporous membranes with alternating positive/negative charge patterns on pore walls effectively mitigate interpore interactions and associated ICP. The heterogeneous charge distribution weakens ion depletion/enrichment overlap, thereby reducing membrane resistance. Notably, the current-enhancing effect becomes more prominent as pore spacing decreases. Bipolar membranes, leveraging their asymmetric charge profiles, suppress ICP at pore entrances and minimize current reduction caused by interpore interaction. Furthermore, both unipolar and bipolar membranes exhibit ion current rectification, with rectification ratios increasing as inter-pore spacing widens. This behavior highlights the interplay between pore arrangement and surface charge heterogeneity in governing nanofluidic transport.

Based on the simulation results, we propose a current prediction formula for porous membranes with different surface charge properties, obtaining the quantitative relationship between optimal current and pore number under various pore spacings. Under smaller pore spacings, ICP being more pronounced, the ionic current exhibits a logarithmic relationship with the number of nanopores, i.e. $I_{total} \sim \ln N_p$. As the pore spacing increases, the pore-pore interactions gradually weaken, and a square-root dependence is observed ($I_{total} \sim \sqrt{N_p}$). When the pore spacing exceeds 1000 nm, the current in porous membranes shows a linear dependence on pore number ($I_{total} \sim N_p$). The revealed mechanism of pore-pore interaction in regulating ionic current through porous membranes provides theoretical guidance for the design of porous membranes.

## Conflict of interest

The authors have no conflicts to disclose.

## Acknowledgment

This research was supported by the National Key R&D Program of China (2023YFF0717105), the Shandong Provincial Natural Science Foundation (ZR2024ME176), the Basic and Applied Basic Research Foundation of Guangdong Province (2025A1515010126), the Shenzhen Science and Technology Program (JCYJ20240813101159005), the National Natural Science Foundation of China (52105579), the Innovation Capability Enhancement Project of Technology-based Small and Medium-sized Enterprises of Shandong Province (2024TSGC0866), and the Key Research and Development Program of Yancheng City (BE2023010).

## Supplementary Information

See supplementary material for simulation details and additional simulation results.